\documentclass[aps,prl,superscriptaddress,twocolumn,final]{revtex4}
\usepackage{amsmath}
\usepackage{graphicx}

\newcommand{\E}{{\mathrm{e}}}
\newcommand{\I}{{\mathrm{i}}}

\begin{document}

\title{Variational finite-difference representation of the
  kinetic energy operator} 
\author{P.~Maragakis}
\affiliation{Division of Engineering and Applied Sciences,  Harvard
  University, 02138 Cambridge MA}

\author{Jos\'e M.\ Soler}
\affiliation{Physics Department, Harvard University, 02138 Cambridge MA}
\affiliation{Departamento de F\'{\i}sica de la Materia Condensada, 
  Universidad Aut\'onoma de Madrid, 28049 Madrid, Spain}

\author{Efthimios Kaxiras}
\affiliation{Division of Engineering and Applied Sciences,  Harvard
  University, 02138 Cambridge MA}
\affiliation{Physics Department, Harvard University, 02138 Cambridge MA}

\begin{abstract}
  A potential disadvantage of real-space-grid electronic structure
  methods is the lack of a variational principle and the concomitant
  increase of total energy with grid refinement.  We show that the
  origin of this feature is the systematic underestimation of the
  kinetic energy by the finite difference representation of the
  Laplacian operator.  We present an alternative representation that
  provides a rigorous upper bound estimate of the true kinetic energy
  and we illustrate its properties with a harmonic oscillator
  potential.  For a more realistic application, we study the
  convergence of the total energy of bulk silicon using a
  real-space-grid density-functional code and employing both the
  conventional and the alternative representations of the kinetic
  energy operator.
\end{abstract}

\maketitle


Electronic structure methods based on finite differences on a real
space grid have gained much support in recent
years~\cite{bernholc:1999:cms,beck:2000:rsm} due to their simplicity
and versatility.  As with plane wave basis sets, their accuracy can be
improved easily and systematically.  In fact, there exists a rigorous
cutoff for the plane waves that can be represented in a given grid,
without aliasing, that provides a convenient connection between the
two schemes.  Soft~\cite{troullier:1991:a} and
ultrasoft~\cite{vanderbilt:1990:u} pseudopotentials, developed in the
plane wave context, can be applied equally well in grid-based methods,
resulting in an accurate and efficient evaluation of the potential
energy.  In contrast with plane waves, the evaluation of the kinetic
energy by finite differences is approximate, but it can be
significantly improved by using high order representations of the
Laplacian
operator~\cite{chelikowsky:1994:hof,chelikowsky:1994:fdp,modine:1997:a}.
However, an important difference between finite difference schemes and
basis set approaches is the lack of a Rayleigh-Ritz variational
principle in the former case.  With finite differences, the accuracy
of the calculation can also be improved systematically by increasing
the grid cutoff (i.e. the grid density).  But denser grids generally
result in {\em higher} energies.  This feature, common to all existing
real-space-grid 
approaches to electronic structure calculations, has been discussed
frequently in the literature (see for example the discussion of
equation (20) in Ref.[\onlinecite{beck:2000:rsm}]).
It is one of the reasons why it is difficult to develop
extrapolation methods and convergence schemes based on minimizing the
total energy.

In this paper we show that the origin of the ``anti variational''
behavior of the total energy in real-space-grid approaches lies in a
systematic underestimation of the kinetic energy by the
finite-difference representation of the Laplacian operator,
independently of the order used.  We propose a simple way to
construct, for any order, an alternative Laplacian representation that
leads to kinetic energies that are higher or equal to the true kinetic
energy.  The paper is organized as follows.  We first briefly present
a way to construct the conventional finite difference representation
of a Laplacian and show that the spectrum of the resulting operator is
lower than the true one.  We then apply a variation of this theme to
construct a representation which has a spectrum higher than the true
one.  We compare the convergence of the one-dimensional harmonic
oscillator energy using the two representations.  Finally we implement
the new representation in a real-space electronic structure code and
examine the convergence properties of the calculations of bulk Si.


We will consider a three-dimensional (3D) Cartesian coordinate system, 
in which the Laplacian is the sum of three one-dimensional (1D)
operators.
For a regular grid in one dimension, a general finite-difference 
expression for the Laplacian of function $\psi(x)$ at point $x_i$ is
\begin{equation}
  \nabla^2 \psi_i = {1 \over a^2} \sum_{j=-N}^N c_j \psi_{i+j},
  \label{laplacian}
\end{equation}
where $\psi_i \equiv \psi(x_i)$.
The constant $1/a^2$ takes care of the dependence on
the grid interval $a$, so that the coefficients $c_j$ 
are independent of it.  Thus, we will use $a=1$ for simplicity in what
follows.

\begin{figure}[htbp]
  \includegraphics[width=\columnwidth]{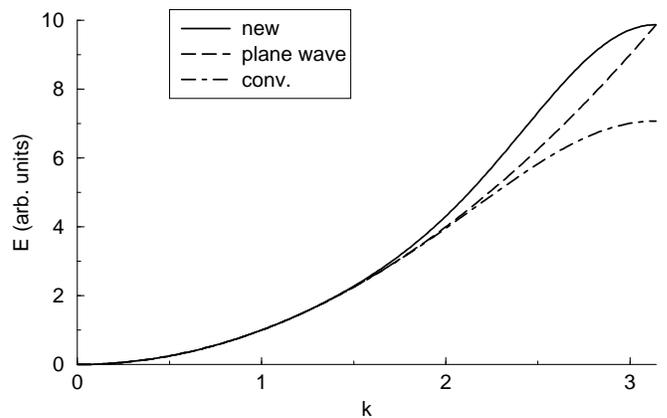}
  \caption{Middle (dashed) line: exact kinetic energy of a plane wave
    $E(k)=k^2$.  Lower (dot-dashed) line: kinetic energy obtained
    using the conventional finite difference Laplacian representation
    of order $N=6$.  Upper (solid) line: energy of the new Laplacian
    representation with $N=6$.}
  \label{fig:kinetic}
\end{figure}

One way to obtain the Laplacian coefficients $c_j$ is through
its eigenvalue equation $E(k) = k^2$, where $E(k)$ is 
(except for a constant factor)
the kinetic energy of a single plane wave:
\begin{equation}
  E(k) = - \E^{-\I kx} \nabla^2 \E^{\I kx} = 
         - c_0 - 2 \sum_{j=1}^{N} c_j \cos{(kj)},
  \label{Ek}
\end{equation}
where we have used $c_{-j}=c_j$ due to the parity of the Laplacian.
Notice that $E(k)$ is periodic by construction, with period $2\pi$,
and that its slope is necessarily zero at the grid's Nyquist limit
$k=\pm \pi$.  Since we have only $N+1$ discrete coefficients to impose
$E(k) = k^2$ in the continuous range $[-\pi, \pi]$, this can be done
only approximately.  In an electronic structure calculation, most of
the spectral weight is concentrated at low values of $k$.  Therefore,
a sensible prescription is to require the Laplacian to be accurate
around $k=0$, by requiring that the value of $E(k)-k^2$ and its $N$
first even derivatives be zero at this point.  The resulting values
of $c_j$ are given in Table I of
Ref.~\onlinecite{chelikowsky:1994:hof}.  However, because of its
periodic character, and its zero slope at $k=\pi$, the resulting
$E(k)$ is a {\it lower bound} of the true kinetic energy $k^2$, as is
shown in Fig.~\ref{fig:kinetic}.  With increasing discretization order
$N$, $E(k)$ approaches $k^2$, but the convergence is always from
below.


A simple variation of the above theme leads to a representation whose
corresponding kinetic energy operator is an upper bound to the true
kinetic energy~\cite{fit:note}.  
Instead of fitting $N$ derivatives of $E(k)$ at $k=0$,
we fit  
only the value and the first $N - 1$ even 
derivatives at $k=0$ and, additionally, the value at $k=\pi$.  
In Table~{\ref{tab:coef_table}} we present the resulting 
coefficients for orders $N=1-6$.

\begin{table}[htbp]
  \begin{center}
    \begin{tabular}[c]{ccccc}
$N$ & $c_i$ & $c_{i+1}$ & $c_{i+2}$ & $c_{i+3}$ \\
\hline
1   & ${\frac{-{{\pi }^2}}{2}}$ 
& $\frac{{{\pi }^2}}{4}$ &$$&$$\\
2   & $-\frac{1}{2}-\frac{3 {\pi}^2}{8}$ & ${\frac{{{\pi }^2}}{4}}$ & 
$\frac{1}{4} - \frac{{\pi}^2}{16}$ &$$\\ 
3   & $-\frac{5}{6}-\frac{5{\pi}^2}{16}  $ & 
$\frac{1}{12} + \frac{15 {\pi}^2}{64} $ &
${\frac{5}{12}} - {\frac{3\,{{\pi }^2}}{32}}$ & 
$ - \frac{1}{12} + \frac{{\pi}^2}{64}$ \\
4   & $ - \frac{77}{72} - \frac{35 {\pi}^2}{128} $ &
${\frac{8}{45}} + {\frac{7\,{{\pi }^2}}{32}}$ & ${\frac{23}{45}} -
{\frac{7\,{{\pi }^2}}{64}}$ & $-{\frac{8}{45}} + {\frac{{{\pi
}^2}}{32}}$ \\
5   & $-{\frac{449}{360}} - {\frac{63\,{{\pi }^2}}{256}}$ &
${\frac{4}{15}} + {\frac{105\,{{\pi }^2}}{512}}$ & ${\frac{59}{105}} -
{\frac{15\,{{\pi }^2}}{128}}$ & $-{\frac{82}{315}} + {\frac{45\,{{\pi
}^2}}{1024}}$ \\
6   & $-{\frac{2497}{1800}} - {\frac{231\,{{\pi }^2}}{1024}}$ &
${\frac{26}{75}} + {\frac{99\,{{\pi }^2}}{512}}$ & ${\frac{493}{840}} -
{\frac{495\,{{\pi }^2}}{4096}}$ & $-{\frac{103}{315}} + {\frac{55\,{{\pi
}^2}}{1024}}$\\ 
\hline
$N$ & $c_{i+4}$ & $c_{i+5}$ & $c_{i+6}$ \\
\hline
4 & ${\frac{17}{720}} - {\frac{{{\pi }^2}}{256}}$\\
5 & ${\frac{311}{5040}} - {\frac{5\,{{\pi }^2}}{512}}$ &
$-{\frac{2}{315}} + {\frac{{{\pi }^2}}{1024}}$ \\
6 & ${\frac{2647}{25200}} - {\frac{33\,{{\pi }^2}}{2048}}$ &
$-{\frac{31}{1575}} + {\frac{3\,{{\pi }^2}}{1024}}$ & ${\frac{1}{600}} -
{\frac{{{\pi }^2}}{4096}}$
    \end{tabular}
    \caption{Laplacian expansion coefficients leading to upper bound
      representation of the kinetic energy operator.}
    \label{tab:coef_table}
  \end{center}
\end{table}

In Fig.~\ref{fig:kinetic} we show the kinetic energy of a plane
wave as a function of its wavevector $k$, in the range $[0, \pi]$.
The middle line is the exact kinetic energy, the lower curve is
the energy of the conventional Laplacian representation of order $N=6$,
and the upper curve is the result of the new Laplacian representation
of order 6.  It is clearly seen that the new representation gives
an upper bound to the true kinetic energy.

\begin{figure}[htbp]
  \begin{center}
 \includegraphics[width=\columnwidth]{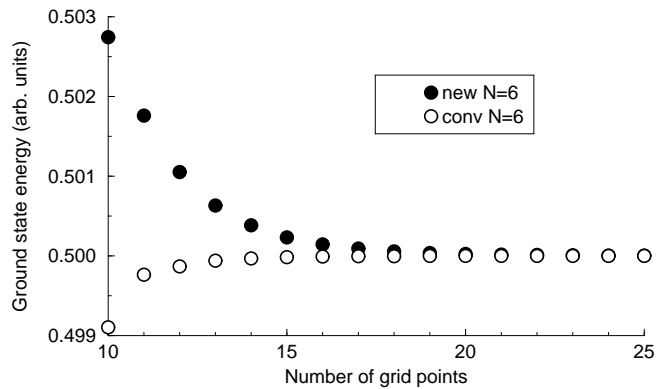}
    \caption{Convergence of the lowest eigenvalue of the harmonic
      oscillator with the number of grid points.  The filled circles
      are the results of the upper bound Laplacian representation,
      the open circles are the results of the conventionalLaplacian.}
    \label{fig:harmonic}
  \end{center}
\end{figure}

The simplest example that clearly demonstrates the properties of the
new Laplacian in action is the 1D harmonic oscillator.
We show the gradual convergence of the calculation using the two
representations by increasing the number of points used to sample the
harmonic oscillator potential $\frac{1}{2} x^2$.  In
Fig.~\ref{fig:harmonic} we plot the lowest eigenvalue, whose
converged value is 0.5, versus the number of points that sample the
interval [-5, 5].  We use a sixth order representation of the
Laplacian for both calculations.  The convergence is faster with
the conventional representation, because it samples better the low 
energy part of the spectrum, but the convergence is from below,
as expected. In contrast, the new representation converges from above 
with increasing hamiltonian size, like a basis-set expansion.

Since the 3D Laplacian in Cartesian coordinates is just a sum of three
independent 1D Laplacians, its representation in a uniform grid has
$6N+1$ nonzero elements in a cross orientation, with $3 c_0$ at the
center and the $N$ elements along the positive and negative sides of
the 3 axes.  We have implemented both the conventional and the upper
bound representation in a real space electronic structure
code~\cite{waghmare:2001:hem}.  We use the new representation in the
calculation of the kinetic energy part of the operator, and the
conventional representation for the solution of the Poisson problem.
The Poisson part of the problem typically converges from above using
conventional real-space Laplacians, since the contribution of the
plane waves enters with a prefactor of $1/G^2$, where $G$ is the
wavevector of the plane wave.  For the same reason one can argue that
the Poisson problem needs a lower discretization in its solution,
since the high energy components are damped by $1/G^2$.

\begin{figure}
  \begin{center}
 \includegraphics[width=\columnwidth]{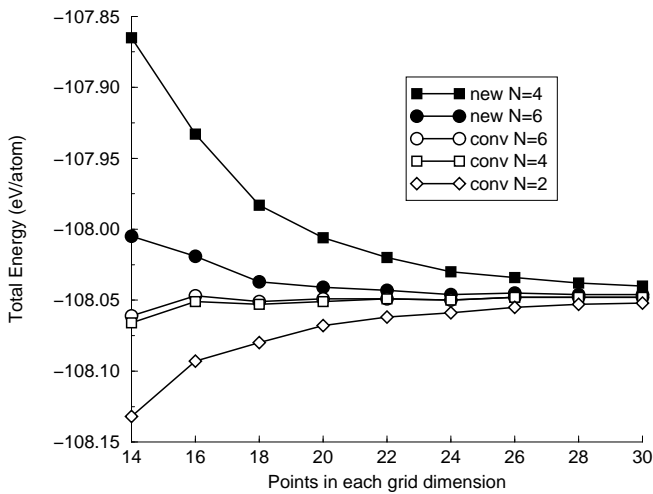}
\caption{Convergence of the total energy calculations for Silicon with
  respect to the number of grid points in each dimension.  The curves,
  from top to bottom correspond to calculations using the new kinetic
  energy formula (filled symbols) with orders 4 (squares), 6
  (circles), and then the conventional formula (open symbols) of order
  6 (circles), 4 (squares), and 2 (diamonds).  The lines are only used
  as a guide to the eye.}
  \label{fig:si-convergence}
  \end{center}
\end{figure}

In Fig.~\ref{fig:si-convergence} we present the convergence
characteristics of a total energy calculation for bulk Si, using
norm-conserving pseudopotentials~\cite{bachelet:1982:pwh}.  An 8 atom
diamond lattice cubic unit cell is used with a $6 \times 6 \times 6$
$k$-point grid in the Monkhorst-Pack scheme~\cite{monkhorst:1976:spb}.
The lattice constant is fixed at 10.264 a.u. and the real-space grid
discretization is gradually increased from 14 to 30 points in each
dimension: this corresponds to an effective plane-wave energy cutoff
varying from 18 to 84 Ryd.  For simplicity, we use the same
order for the Laplacian in the kinetic and Hartree energies. The
lowest curve (left pointing triangles) shows the results of the second
order Laplacian that is used in HARES~\cite{modine:1997:a}.  The total
energy converges from below, implying that the major source of error
comes from the kinetic energy operator.  The two curves immediately
above it (up-pointing triangles and rhombuses) show the total energy
convergence using the conventional formulas of order 4 and 6.  The
energy converges from below and quickly reaches its correct value.
This is partially due to a cancellation of errors between the kinetic
and Hartree energies, but the convergence is not necessarily monotonic
as is manifest at 16 and 18 grid points per dimension.  The uppermost
curve (circles) shows the results using the new Laplacian of order 4
for the kinetic energy, and the conventional Laplacian for the Hartree
energy.  The total energy convergence is slower, similar to the second
order conventional representation, because there is no error
cancellation, since both the kinetic and Hartree energies converge
from above. Also, fewer terms are used to approximate the low-$k$
region, while fixing the Laplacian value at $k=\pi$ distorts
considerably the kinetic energy eigenspectrum of
Fig.~\ref{fig:kinetic} at low order expansions.  Finally, the curve
immediately below it (squares) shows the total energy convergence for
the new representation of order 6, which has a very satisfactory
numerical converge from above.


In summary, we derived a new finite difference representation of the
Laplacian that is guaranteed to give an upper bound value for the
kinetic energy operator.  We demonstrated this basic property of the
1D Laplacian through the harmonic oscillator potential and implemented
and tested the 3D version in a real-space electronic structure code.
We find that for sufficiently high order of the Laplacian, which for
the case of bulk Si is 6, the new form exhibits very satisfactory
numerical convergence of the total energy, while establishing the
variational character of the real space method.


PM acknowledges the hospitality of Dr.~L.~A.~A.~Nikolopoulos at the
Institute for Electronic Structure and Lasers, Heraklion, Greece, in
July 2000, and the support of the European Commission through
TMR grant number ERB FMGE CT950051 (the TRACS Programme at EPCC)".  
JMS acknowledges support from Fundaci\'on Areces and from
Spain's MEC grant BFM2000-1312.  

\bibliographystyle{prsty}
\bibliography{laplacian}

\begin{thebibliography}{11}
\bibitem[{fit()}]{fit:note}
 (????), \bibinfo{note}{another possibility, not explored here, would be to fix
  some of the coefficients by a least square fit. Although the variational
  character would not be guaranteed, it might well result in practice.}
\end{thebibliography}


\end{document}